# Exploration of novel ICP using helicon antennas with zero magnetic field

Ye Tao[1,2], Lei Chang[1†], Dingzhou Li[1], Yingxin Zhao[1]

(1.School of Electrical Engineering, Chongqing University, Chongqing 400044, China;

2.School of Artificial Intelligence, Chongqing University of Technology, Chongqing 401135, China)

†leichang@cqu.edu.cn

**Abstract**：Inductively coupled plasma (ICP) attracts great attention from aspects of fundamental research and practical applications, and efficient power coupling is highly desirable for both of them. The present study explores a novel strategy for efficient ICP through using helicon antennas with zero external magnetic field. Specific research is devoted to the effects of antenna geometry (loop, half-helix, Boswell, Nagoya III), driving frequency (13.56–54.24 MHz) and radial density profile (Gaussian and parabolic) on power coupling. Findings reveal that: loop antenna yields higher power deposition efficiency than half-helix, Boswell, and Nagoya III antennas, driving frequency gives negligible effects, and parabolic density profile results in more efficient power coupling than Gaussian density profile especially in the radial direction, for the conditions employed here. Therefore, it is suggested that for this novel ICP strategy one should use loop antenna with parabolic density profile, and the industrial frequency of 13.56 MHz can work well. This study provides a valuable reference for the novel design of efficient ICP sources, which could be used for material processing and space propulsion, etc.

**Key words**：Inductively coupled plasma; Antenna Geometry; Power Deposition; Driving Frequency

## I. INTRODUCTION

Inductively coupled plasma (ICP) attracts great attention from aspects of fundamental research and practical applications due to its unique ability to generate high-density, low-pressure plasma with exceptional uniformity and control[1], [2]. In fundamental research, ICP serves as a versatile platform for studying plasma physics, including wave-particle interactions[3], [4], plasma instabilities[5], [6], [7], and electron energy distributions[8], offering critical insights into the behavior of ionized gases under diverse conditions. Its high ionization efficiency and stable plasma characteristics enable detailed investigations of atomic and molecular processes[9], such as excitation, dissociation, and recombination, which are essential for refining theoretical models in plasma science[10]. In practical applications, efficient power coupling is highly desirable[11], [12], [13], [14], [15]. ICP is a cornerstone in advanced material processing and electric propulsion. In material processing, ICP is widely adopted for large-scale[16], low-cost[17], and highly uniform processes, such as etching[18] and deposition[19] in semiconductor manufacturing and the fabrication of thin films and nanostructures[20]. Its ability to maintain uniform plasma over large areas ensures consistent material quality, reducing production costs and enabling applications in industries like microelectronics and renewable energy. In electric propulsion, ICP is integral to the development of high-power[21], long-lifetime[22], and highly reliable systems[23], particularly in electrode-less RF plasma thrusters like helicon thrusters[24], [25]. These systems leverage ICP's efficient power coupling to generate high-density plasma, enabling robust thrust performance for extended space missions. The combination of high power output, operational longevity, and reliability makes ICP-based thrusters a promising technology for future deep-space exploration, where precise control and efficiency are paramount. Overall, both fundamental research and practical applications of ICP requires highly efficient strategy.

The present study explores a novel strategy for efficient ICP through using helicon antennas with zero external magnetic field. Existing ICP sources utilizes simple antenna geometry, e.g. solenoid. Helicon plasma employs various antennas but for non-zero magnetic field. Here, we combine them together: to explore a novel ICP strategy using helicon antennas with zero magnetic field.

The structure of this article is as follows. Section II introduces the theoretical model and computational scheme. Section III is the description of the results and analysis. Section IV is the conclusion.

## II. THEORETICAL MODEL AND COMPUTATIONAL SCHEME

This work primarily utilizes the HELIC program for calculations. The HELIC program was mainly developed by D. Arnush and F. F. Chen [26], [27], [28], [29]. Its fundamental theoretical model consists of Maxwell's equations for radially non-uniform plasmas with a standard cold plasma medium tensor, which can be transformed into the following coupled differential equations through Fourier transformation:

$$\frac{\partial E_\varphi}{\partial r} = \frac{im}{r} E_r - \frac{E_\varphi}{r} + i\omega B_z \tag{1}$$

$$\frac{\partial E_z}{\partial r} = ikE_r - i\omega B_\varphi \tag{2}$$

$$\frac{\partial B_\varphi}{\partial r} = \frac{m}{r}\frac{k}{\omega} E_\varphi - \frac{iB_\varphi}{r} + (P - \frac{m^2}{k_0^2 r^2})\frac{\omega}{c^2} E_z \tag{3}$$

$$i\frac{\partial B_z}{\partial r} = -\frac{\omega}{c^2} iDE_r + (k^2 - k_0^2 S)\frac{E_\varphi}{\omega} + \frac{m}{r}\frac{k}{\omega} E_z \tag{4}$$

Here, a cylindrical coordinate system $(r, \varphi, z)$ and first-order perturbation form of $e^{i(m\varphi+kz-\omega t)}$ have been chosen, where $m$ is the azimuthal mode number, $k$ is the axial mode number, $\omega$ is the wave frequency, E and B denote the wave electric and magnetic fields, respectively, $k_0 = \omega/c$ and $c$ is the speed of light. Additionally, the cold plasma medium parameters S, D, and P, following Stix notation, can be expressed as:

$$R = 1 - \sum_s \frac{\omega_{ps}^2}{\omega(\omega+iv_s)} \left[ \frac{(\omega+iv_s)}{(\omega+iv_s)-\omega_{cs}} \right] \tag{5}$$

$$L = 1 - \sum_s \frac{\omega_{ps}^2}{\omega(\omega+iv_s)} \left[ \frac{(\omega+iv_s)}{(\omega+iv_s)-\omega_{cs}} \right] \tag{6}$$

$$S = \frac{1}{2}(R+L), \ D = \frac{1}{2}(R-L), \ P = 1 - \sum_s \frac{\omega_{ps}^2}{\omega(\omega+iv_s)} \tag{7}$$

Here, the phenomenological collision rate $v$ is introduced to solve for electron-neutral and electron-ion collisions. $\omega_{ps} = \sqrt{q_s^2 n_s / (\varepsilon_0 m_s)}$ is the plasma frequency, $\omega_{cs} = q_{cs} B_0 / m_s$ is the electron cyclotron frequency, $q_s$ is the charge, $m_s$ is the particle mass, and $n_s$ is the plasma density. The subscripts $s$ and 0 denote the plasma species (electron and ion) and the equilibrium value, respectively. The parameters R and L represent the right- and left-handed circular polarization modes, respectively. These expressions are fully general and can accommodate multiple ion species. In this paper, however, we assume that the ions in the plasma have a single charge, meaning there is only one type of ion.

In this work, the radial distribution Pr and axial distribution Pz of plasma power deposition were calculated using the HELIC program[30], [31]. The primary input parameters for the program include the plasma radius a=5cm, antenna radius b=5.5cm. The length is assumed to be infinite, although the axial range of 0.63 < z < 0.63 m is computed. Plasma density $n_0$ uniformly distributed in the range of $10^{15}$ cm$^{-3}$ to $10^{18}$ cm$^{-3}$, a uniform background DC magnetic field strength $B_0=10^{-4}$ G, and radio frequency (RF) source frequencies of 13.56 MHz, 27.12 MHz, 40.68 MHz, and 54.24 MHz. The electron temperature was set to 3 eV, and the working pressure was 0.133322 Pa. The wave number range 5<k<50 was selected based on the spectrum of most helical Ar discharges. This study investigates four types of RF antennas—half-helix, Boswell, Nagoya III (for exciting the m=1 mode), and loop (for m=0 mode)—all with identical lengths of 0.2 m. The simulation employ two commonly observed plasma density profiles: a parabolic and a Gaussian radial distribution. As illustrated in Figure 1, both profiles exhibit identical peak densities on-axis but differ significantly in their radial gradients. The edge densities of these profiles are set to 10% and 1% of the peak value, respectively. The choice of these distinct density profiles is critical, as the radial density gradient and edge density profoundly influence non-resonant mode conversion (from helicon waves to Trivelpiece-Gould waves) and power absorption. By comparing these two profiles with markedly different gradients, the study aims to elucidate their respective impacts on plasma dynamics. According to the constructed RF micro discharge plasma model, the HELIC program computed the corresponding power spectral functions, which were then analyzed and compared.

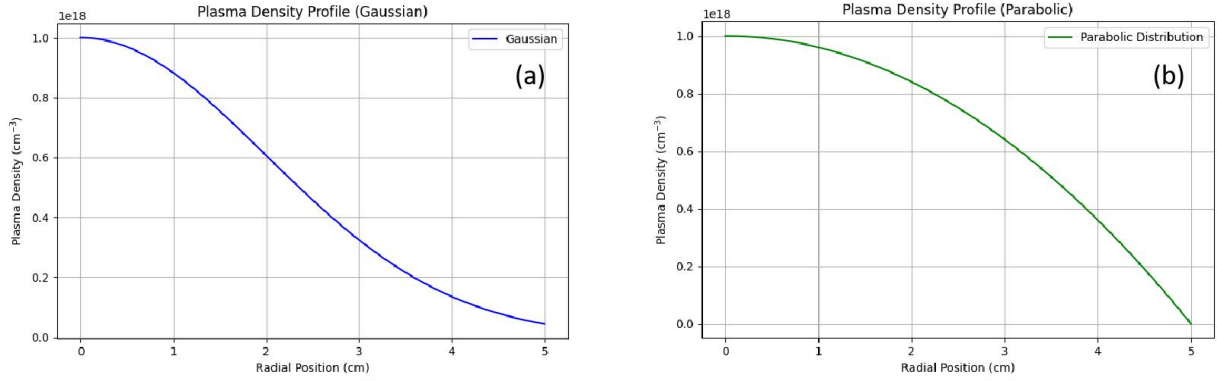

Fig. 1. Radial profiles of plasma density: (a) Gaussian, (b) Parabolic.

## III. RESULT AND ANALYSIS

Figure 2 presents the axial power deposition profiles for four representative antennas—loop, half-helix, Nagoya III, and Boswell—operating at 13.56 MHz, under a Gaussian plasma density distribution. The power deposition for each antenna peaks at the axial center ($z = 0$), consistent with the Gaussian profile, where electron density is maximized at the center and decreases rapidly toward the edges. This distribution results in highly localized power deposition at the axial midpoint.

The peak power deposition values rank as follows: loop > half-helix > Nagoya III > Boswell. This ordering reflects the distinct design characteristics and performance efficiencies of the antennas. The loop antenna, benefiting from superior radiation efficiency and optimized current distribution, achieves the highest peak power deposition due to its enhanced energy concentration. The half-helix antenna, with its spiral geometry, provides strong directional radiation, securing the second-highest power deposition. Conversely, the Nagoya III and Boswell antennas exhibit lower peak values, likely due to their more intricate designs or reduced radiation efficiency.

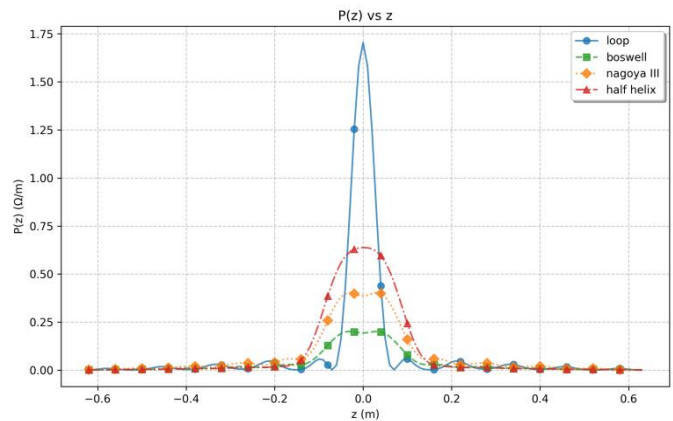

Fig. 2. Comparison of axial power deposition for different antenna geometry at 13.56MHz under Gaussian density distribution.

The influence of varying RF driving frequencies on the axial power deposition of antennas was examined, with a specific focus on the performance of Loop antennas, as depicted in Figure 3. The results indicate that increasing the driving frequency from 13.56 MHz to 27.12 MHz leads to a significant enhancement in power deposition. However, further increases to 40.68 MHz and 54.24 MHz yield negligible changes in power deposition. This behavior can be attributed to three primary factors: antenna operational frequency band limitations, impedance matching efficiency, and plasma response characteristics.

At the transition from 13.56 MHz to 27.12 MHz, the higher frequency likely improves resonance and coupling efficiency, resulting in greater power deposition. Beyond this range, at 40.68 MHz and 54.24 MHz, the antenna and plasma system approach the limits of impedance matching, leading to a saturation of power deposition. This trend is consistent across other antenna types evaluated in the study.

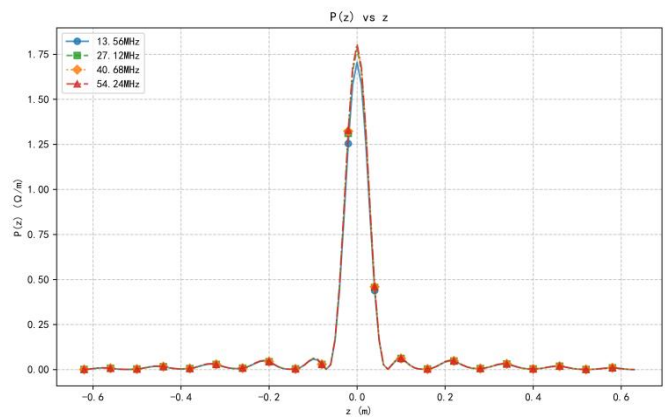

Fig. 3. Axial power deposition of loop antennas as a function of RF driving frequency under Gaussian density distribution.

To investigate the influence of plasma density distribution on antenna power deposition, we analyzed and compared the axial power deposition profiles of four representative antenna types under a parabolic plasma density distribution at various driving frequencies.

We first evaluated the axial power deposition of different antenna types at a driving frequency of 13.56 MHz, as shown in Figure 4. The results indicate that, under a parabolic density profile, the peak positions and waveform patterns of axial power deposition across antenna types at different frequencies are similar to those observed under a Gaussian profile. However, the peak power deposition values under the parabolic distribution are consistently higher than those under the Gaussian distribution. This difference likel y stems from the parabolic density profile's gradual decrease from the axis to the edge, which facilitates more uniform radial energy diffusion, reduces axial saturation, and enhances wave penetration depth, there by improving overall power absorption. These observations are consistent with previous research findings.

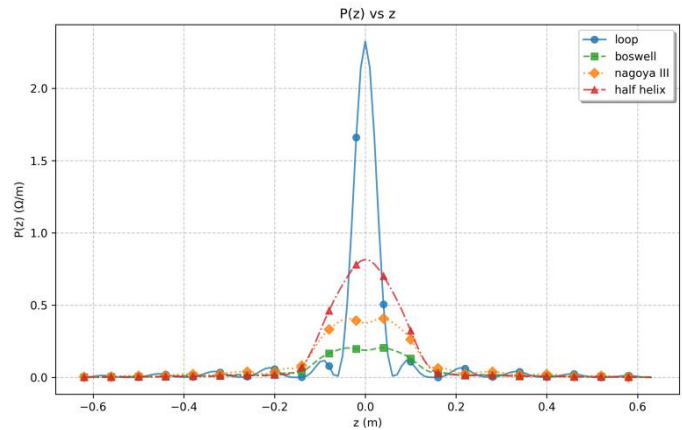

Fig. 4. Comparison of axial power deposition for different antenna types under 13.56MHz frequencies in a parabolic density profile.

Regarding the effect of RF driving frequencies on axial power deposition, antennas operating under a parabolic density profile exhibit trends in power deposition variation across frequencies that are consistent with those observed under a Gaussian profile. Taking the half-helix antenna as an example, as illustrated in Figure 5, we observed that increasing the driving frequency from 13.56 MHz to 27.12 MHz results in increased power deposition. Further increases to 40.68 MHz and 54.24 MHz lead to stabilized power deposition levels across all antenna configurations. Similar trends were observed for the other antenna types.

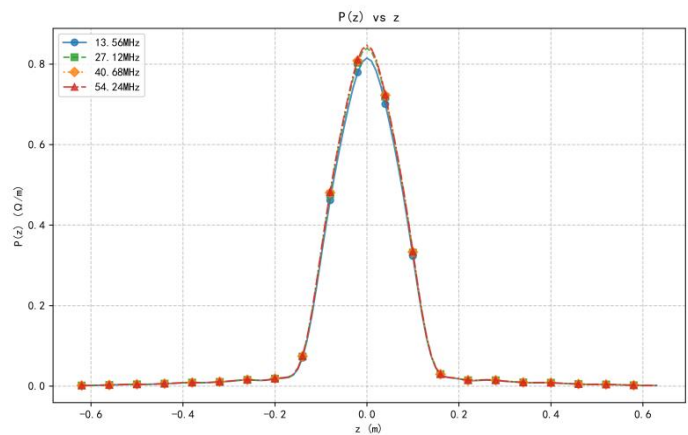

Fig. 5. Frequency-dependent axial power deposition of half-helix antenna under parabolic density.

To further explore the effects of plasma density profiles and RF frequencies on antenna performance, we conducted a comparative analysis of radial power deposition for four representative antenna types. Initially, we investigated the radial power deposition at a driving frequency of 13.56 MHz under a Gaussian plasma density distribution, as illustrated in Figure 6. The results reveal that radial power deposition increases with radial distance r, peaking at approximately r=0.037m (roughly 0.7 times the plasma radius), before declining toward the plasma edge. This behavior is primarily driven by the interaction between the radial distribution of the plasma frequency ($\omega_p$) and the incident electromagnetic wave frequency ($\omega$), coupled with the propagation characteristics of electromagnetic waves in a non-uniform plasma.

The observed trend can be explained as follows: Near the plasma center (r is small), the high plasma frequency ($\omega_p$) prevents electromagnetic wave propagation, creating a cut-off region where waves are partially reflected or weakly absorbed, resulting in low power deposition. As r increases, $\omega_p$ decreases, and in the region where $\omega_p \approx \omega$, resonant absorption occurs, efficiently converting electromagnetic energy into plasma thermal energy and leading to a peak in power deposition at approximately r≈0.7R. Beyond this point, as r continues to increase, electromagnetic waves propagate more freely, but absorption diminishes, causing a decline in power deposition.

The analysis also indicates that antenna type influences coupling efficiency by altering the wave mode without changing the overall radial trend. The coupling efficiencies rank as follows: half-helix > loop > Nagoya III > Boswell. Notably, the half-helix and loop antennas exhibit comparable and significantly higher power deposition than the Nagoya III and Boswell antennas. This difference is likely due to variations in antenna design and electromagnetic properties. The half-helix and loop antennas likely benefit from more efficient current distributions and stronger coupling with the surrounding plasma, enhancing energy transfer in the radial direction. Their geometric configurations may amplify electric field intensity and current flow along the antenna structure, resulting in greater power deposition.

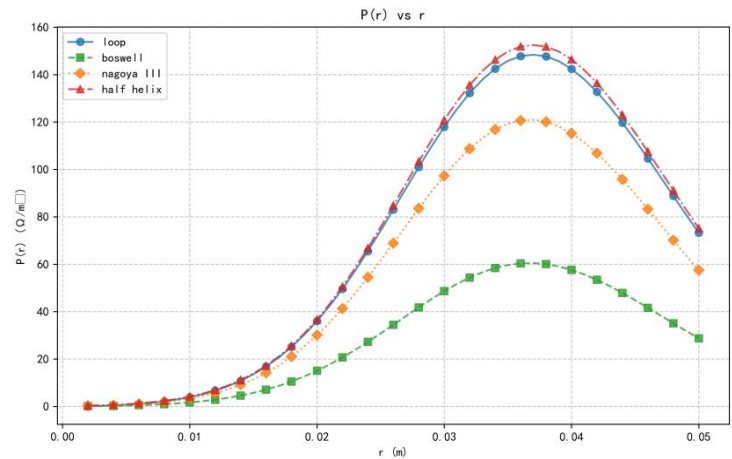

Fig. 6. Comparison of radial power deposition for different antenna types under 13.56MHz frequencies in a Gaussian density profile.

To assess the impact of varying RF driving frequencies on radial power deposition under a Gaussian plasma density distribution, we analyzed the trends in radial power deposition across different antenna types. These trends closely mirror those observed in axial power deposition under equivalent conditions. Taking the loop antenna as an example, as shown in Figure 7, we observed that increasing the driving frequency from 13.56 MHz to 27.12 MHz results in a general increase in radial power deposition. However, further frequency increases to 40.68 MHz and 54.24 MHz yield negligible changes in power deposition. This behavior suggests that the factors governing axial power deposition—such as antenna operating frequency limitations, impedance matching efficiency, and plasma response characteristics—also influence radial power deposition.

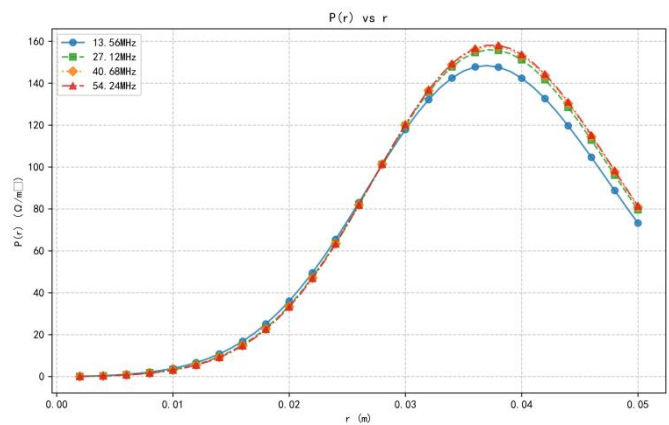

Fig. 7. Effect of different frequencies on radial power deposition of loop antennas under Gaussian distribution.

We conducted a comparative analysis of the radial power deposition for four representative antenna types at a driving frequency of 13.56 MHz under a parabolic plasma density distribution, as illustrated in Figure 8. The radial power deposition curves for these antennas exhibit similarities to those observed under a Gaussian distribution, but with notably higher peak values and peak positions shifted closer to the plasma edge. Specifically, the power deposition increases with radial distance r, reaching a maximum at approximately r=0.045m (roughly 0.9 times the plasma radius), before declining toward the edge. Compared to the Gaussian distribution, the decay rate under the parabolic profile is more pronounced.

This behavior can be attributed to the parabolic distribution's steep density gradient at the plasma core, which induces stronger wave reflection. As a result, the peak power deposition is higher, and the deposition profile is narrower, with the peak shifted toward the edge. In the low-density outer regions, the remaining wave energy diminishes rapidly, leading to a steeper decay in power deposition compared to the Gaussian profile.

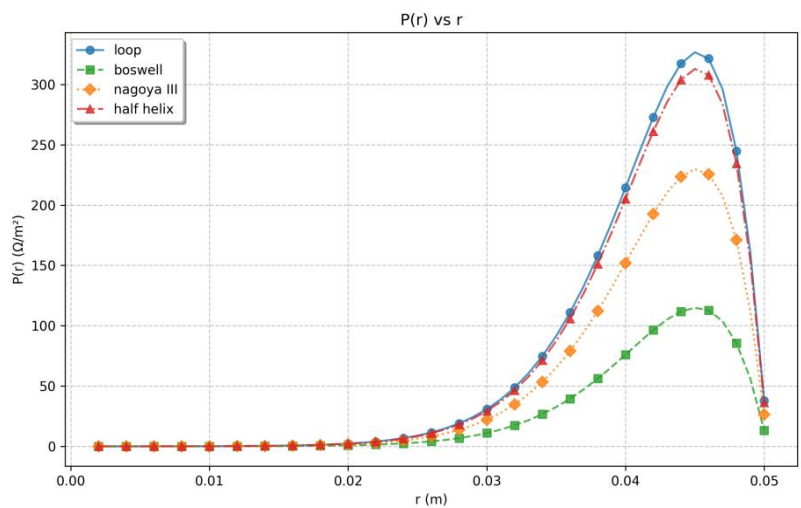

Fig. 8. Comparison of radial power deposition for various antenna types under 13.56MHz frequencies in a parabolic plasma density distribution.

The radial power deposition characteristics of various antennas under parabolic plasma density distribution exhibit similar trends to those observed in Gaussian distributions, take the loop antenna as an example illustrated in Figure 9. Notably, at a driving frequency of 54.24 MHz, all antenna types demonstrate approximately double the peak power deposition compared to Gaussian density conditions.

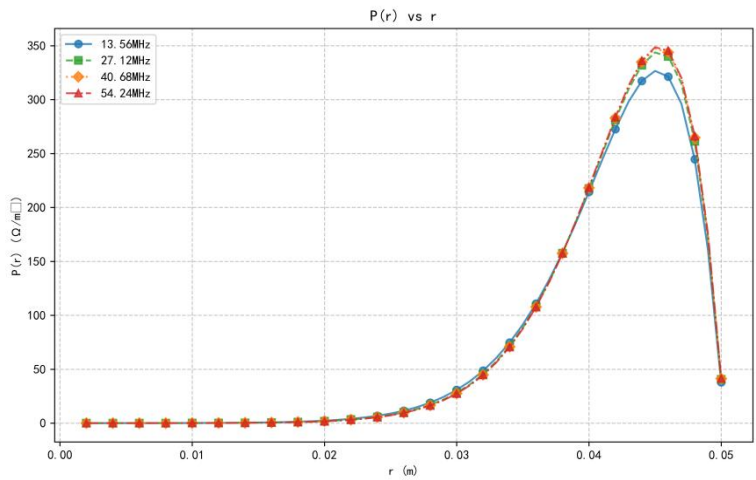

Fig. 9. Effect of different frequencies on radial power deposition for loop antenna under a parabolic distribution.

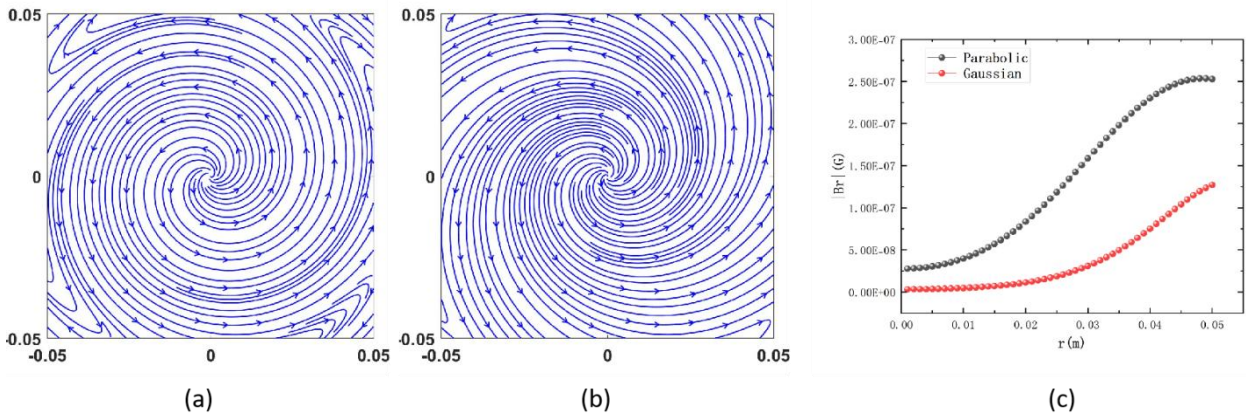

Fig. 10. Perpendicular structures of half-helix antenna wave magnetic field for 13.56MHz driving frequencies: (a) parabolic density profiles; (b) Gaussian density profiles　(c) Comparison between Parabola and Gaussian.

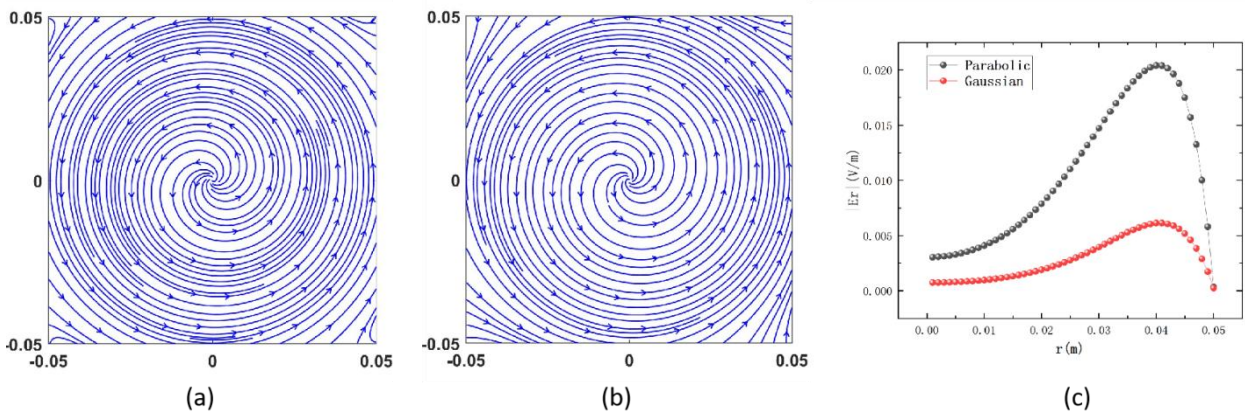

Fig. 11. Perpendicular structures of half-helix antenna wave electric field for 13.56MHz driving frequencies: (a) parabolic density profiles; (b) Gaussian density profiles　(c) Comparison between Parabola and Gaussian.

Next, we analyzed the impact of different density distribution on the wave magnetic field and wave electric field of plasma excited by the antenna, taking the half-helix antenna at 13.56MHz as an example, as shown in Figures 10 and 11. From the figures, it can be observed that the inward spiral streamlines of both the wave magnetic field and wave electric field are counterclockwise, with the spiral becoming more dispersed from the center to the edges. This trend is more pronounced in the Br plots. The spiral under the parabolic density is more uniform, with a wider distribution of lines and a more balanced intensity between the core and edges; In contrast, the spiral under the Gaussian density is more centralized, with a higher line density at the core and sparser lines at the periphery. This trend is more evident in the comparison chart of Br and Er. This could also be a reason why the parabolic density distribution results in higher energy deposition. Similar trends are observed across other radio frequency ranges.

To further investigate the effect of different driving frequencies on antenna power deposition, we analyzed the variation of the total load resistance of each antenna with driving frequency, as shown in Figure 12. From the figure, it can be observed that when the driving frequency increases from 13.56 MHz to 27.12 MHz, the total load resistance of all antenna types exhibits a significant increase. However, as the driving frequency continues to rise, the total load resistance of each antenna type remains essentially unchanged. This trend aligns with the behavior observed in power deposition.

Additionally, the results indicate that the total load resistance under a parabolic density distribution is greater than that under a Gaussian density distribution, further confirming the conclusion drawn from power deposition analysis. Furthermore, the figure shows that the Loop antenna has the highest total load resistance, followed by the half-helix, then the Nagoya III antenna, while the Boswell antenna exhibits the lowest total load resistance. This observation is consistent with the previous findings on power deposition.

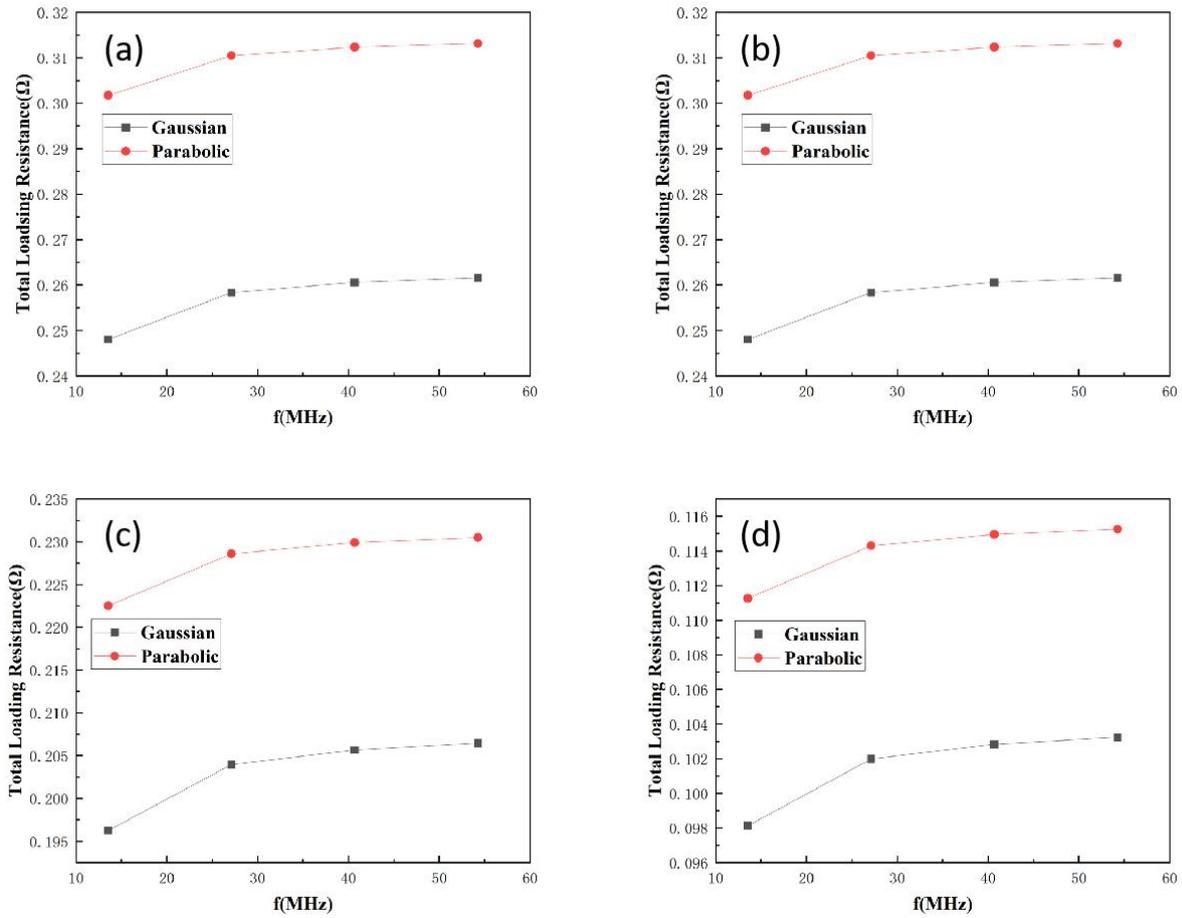

Fig. 12. Total load resistance of various antennas as a function of RF frequency:

(a) Loop, (b) half-helix, (c) Nagoya III, (d) Boswell.

## IV. CONCLUSION

This study introduces a novel approach to enhancing inductively coupled plasma (ICP) efficiency using helicon antennas in the absence of an external magnetic field. Utilizing the HELIC program, we investigated the effects of antenna geometry (loop, half-helix, Boswell, and Nagoya III), driving frequency (13.56–54.24 MHz), and radial plasma density profiles (Gaussian and parabolic) on power coupling efficiency. The results demonstrate that the loop antenna exhibits significantly higher axial power deposition efficiency compared to the half-helix, Boswell, and Nagoya III antennas.

Regarding the influence of driving frequency, increasing the frequency shifts the power deposition toward the plasma edge, resulting in an overall increase in power deposition. However, this enhancement plateaus beyond 27.12 MHz. With respect to the plasma density profile, the parabolic distribution yields superior radial power deposition compared to the Gaussian distribution, with peak values approximately twice as high.

Based on these findings, subsequent experimental validation will focus on the loop antenna operated without an external magnetic field to further optimize power coupling efficiency.

ACKNOWLEDGMENTS

This research is supported by National Natural Science Foundation of China (92271113, 12411540222, 12481540165), the Fundamental Research Funds for Central Universities (2022CDJQY-003), the Chongqing Entrepreneurship and Innovation Support Programme for Overseas Returnees (CX2022004), and the Natural Science Foundation Project of Chongqing (CSTB2025NSCQ-GPX0725).